# Explaining Bright Radar Reflections Below The Martian South Polar Layered Deposits Without Liquid Water

D. E. Lalich[1]*, A. G. Hayes[1], and V. Poggiali[1], [1]Cornell Center for Astrophysics and Space Science, *Corresponding Author: dlalich@astro.cornell.edu

## Introduction

*Recent discoveries of anomalously bright radar reflections below the Mars South Polar Layered Deposit (SPLD) have sparked new speculation that liquid water may be present below the ice cap. The reflections, discovered in data acquired by the Mars Advanced Radar for Subsurface and Ionospheric Sounding (MARSIS) on board the Mars Express orbiter, were interpreted as reflections from damp materials or even subsurface ponds and lakes similar to those found beneath Earth's ice sheets. Recent studies, however, have questioned the feasibility of melting and maintaining liquid water below the SPLD. Herein, we compare radar simulations to MARSIS observations in order to present an alternate hypothesis: that the bright reflections are the result of interference between multiple layer boundaries, with no liquid water present. This new interpretation is more consistent with known conditions on modern Mars.*

The SPLD is a kilometers-thick formation of relatively pure water ice that takes its name from the ubiquitous layering visible in both optical imagery and radar sounding data [1]. Recently, radar observations collected by MARSIS have revealed strong reflections at the base of the deposit in an isolated region centered at $193^0$E, $81^0$S [2]. These reflections are far too powerful to be the result of a simple contact between SPLD ice and the underlying bedrock. Instead, they were interpreted as strong reflections caused by liquid water [2, 3, 4]. Lakes and other hydrological features are common beneath ice sheets on Earth [5, 6, 7], but so far there has been no evidence of liquid water on present day Mars near the poles.

While the possibility of Martian subglacial lakes is exciting, recent studies have questioned the feasibility of that interpretation. Thermal modeling has shown that without a heat source akin to a recently emplaced magma chamber beneath the surface, the basal temperatures required to melt ice are difficult to reach, even for brines [8, 9]. There is no independent evidence for such a heat source as of yet. In addition, the location of the bright reflections does not seem to match any of the potential lake locations predicted by the hydraulic potential beneath the SPLD [10].

In light of these inconsistencies, it is necessary to consider alternate hypotheses for the observed radar returns. Previous work involving data from the Shallow Radar (SHARAD) instrument on board the Mars Reconnaissance Orbiter (MRO) has shown that radar reflections in layered deposits can be greatly affected by constructive and destructive interference [11, 12, 13]. Under the right conditions, such interferences can create reflections that are much brighter than might otherwise be expected. Here we use a one-dimensional radar sounding model to show that strong reflections consistent with those observed by MARSIS beneath the SPLD can be generated without the need for liquid water, using only materials already known to be present in the polar cap. Our simulations produce a strong match to observed waveforms and basal echo powers while remaining consistent with the previously inferred conditions at the base of the SPLD.

*Radar Sounder Simulation Description*

In order to simulate MARSIS echoes for a variety of subsurface layering scenarios (Figure 1) we apply the same basic one-dimensional modeling procedure commonly used to interpret MARSIS and SHARAD observations [2, 3, 14] (see methods). First, we calculate the effective Fresnel reflection coefficient for a given model stratigraphy at each frequency sampled by MARSIS. Next, we multiply the reflection coefficient by a synthetic "chirp" approximating the transmitted MARSIS signal in frequency space in order to determine the radar response. Standard pulse-compression processing is then performed by multiplying the simulated radar response by the complex conjugate of the chirp. Applying an inverse fast-Fourier transform brings the resulting signal into the time domain, where it represents a single radar return (Figure 2). In keeping with previous work, we do not simulate Doppler effects and assume that any Doppler processing has a negligible effect on relative reflection power [2, 3, 14]. While MARSIS commonly operates in subsurface sounding mode at three different center frequencies (3, 4, and 5 MHz), the recent studies claiming detection of liquid water focused predominantly on the 4 MHz data, as it is the most abundant [2, 3, 4]. Because interference patterns are heavily dependent on wavelength, however, we simulated each layer configuration using all three frequency modes. The strong secondary peaks visible in the simulated echoes are a processing artifact known as "side lobes." Side lobes were filtered out of the observed data, but as the filtering process does not change the relative power of the surface and subsurface echoes we did not implement it for our simulations.

There are few constraints on layer thickness or composition near the base of the SPLD, making it virtually impossible to model all potential stratigraphic scenarios. Instead, we focus on a small number of plausible configurations to explore the range of radar reflections that may be produced (Figure 1). For our simulations, layers are composed of four materials: free space, water ice, $CO_2$ ice, and basalt. In order to calculate the effective reflection coefficient, we must assign each layer a corresponding permittivity, which is an intrinsic property of the material describing its interaction with electromagnetic radiation passing through it. The permittivity of each material was either taken from laboratory measurements or calculated using established formulas (see methods).

In each simulation, the first layer is a semi-infinite half-space representing the Martian atmosphere and is assigned a permittivity equal to that of free space. Each simulation also includes a 1.4 km thick layer of mostly pure water ice with a basaltic dust content of 10% by volume, representing the SPLD material between the surface and basal reflections. In actuality the SPLD is made up of many internal layers of varying compositions [15, 16, 17], but as we are only interested in the surface and basal reflections, we simply adopt the bulk composition inferred by Plaut et al. [18]. The thickness of this icy layer is approximately 1.4 km, consistent with previous estimates of SPLD thickness in the region containing the bright reflectors [2]. The final layer that is consistent across all simulations is the lowermost bedrock layer, which is a semi-infinite half space composed of pure basaltic rock.

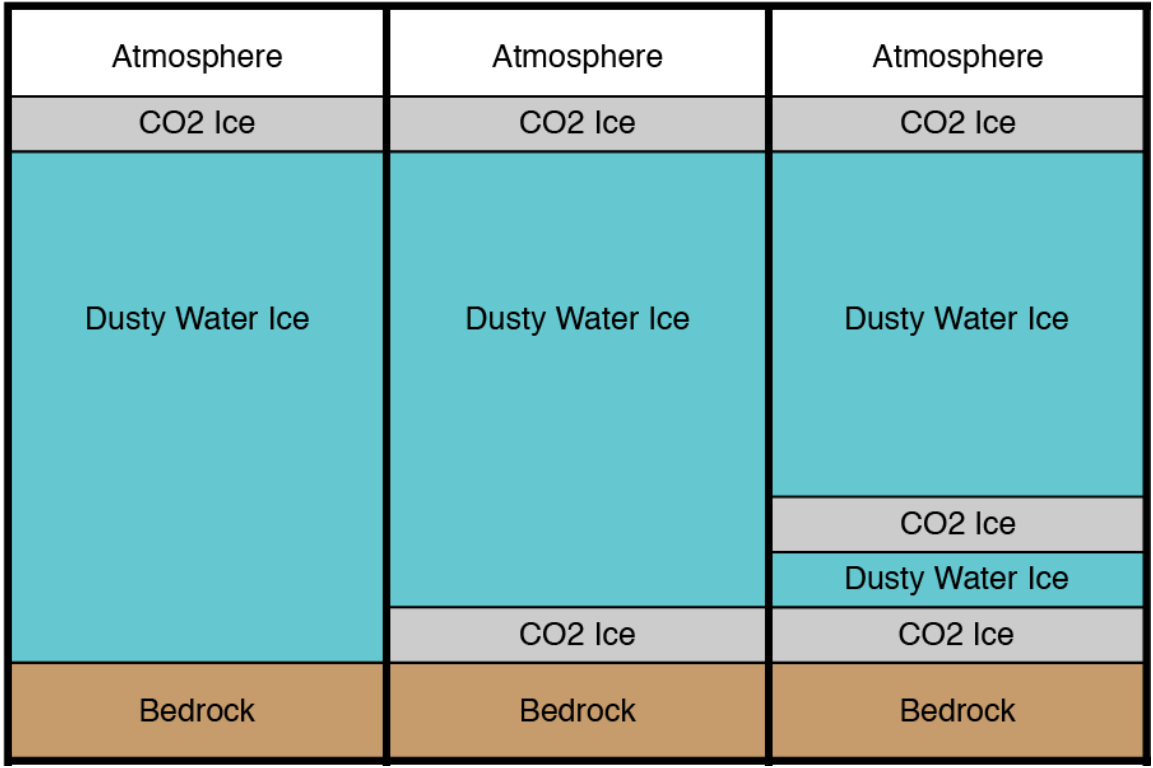

*Figure 1: Schematic diagrams of the three scenarios simulated for this study. Simulations were run both with and without the surface CO2 ice for each scenario. Note that layers are not to scale.*

We simulate three different scenarios at the base of the SPLD. In the first scenario, we add no additional layers between the main deposit of dusty ice and the bedrock. This first scenario does not produce bright reflections and is mainly provided as a point of comparison for our other simulations. For the next scenario, we insert a single layer of pure $CO_2$ ice between the main deposit and the bedrock, and vary its thickness from 1-100 m. The third scenario consists of two $CO_2$ ice layers separated by a dusty water ice layer at the base of the main deposit. The thickness of each of these three subsurface layers is allowed to vary in concert between 1-100 m (all three layers are always equal in thickness). As a special case, we ran one simulation with a "wet" basal surface (see figure 2).

In addition to different basal layering scenarios, we also investigate the effect of the South Polar Residual Ice Cap (SPRIC). The SPRIC is composed of solid $CO_2$ ice up to a few meters thick [19]. While MARSIS cannot resolve such a thin layer, the $CO_2$ ice can still alter the surface reflection power and allow more transmitted energy to penetrate into the deposit, potentially resulting in stronger subsurface reflections. To explore how the presence of the SPRIC impacts basal reflectivity, we ran simulations for each scenario described above both with and without a 2 m thick layer of CO2 ice inserted at the surface. While the SPRIC can be thicker in places, it is often excavated by pits or other sublimation features [1]. We chose to use a thickness consistent with the base of these pits.

Our simulated waveforms bear a strong qualitative resemblance to the observed MARSIS waveforms (figure 2). There is no obvious distinction between simulation results for multi-layer scenarios and for the "wet" scenario proposed in previous work. Thus we should not expect to be able to differentiate between the two hypotheses on the basis of echo shape alone.

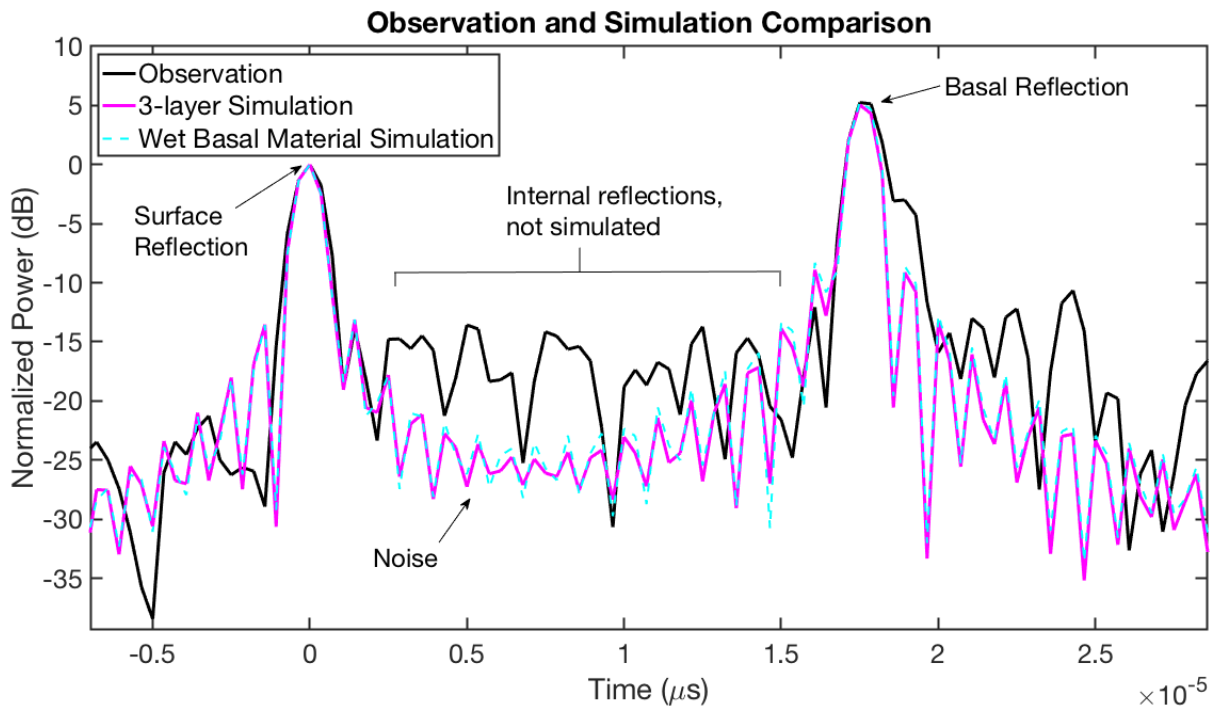

*Figure 2: Comparison between an observed (black) and simulated (pink, cyan) MARSIS echoes. The observed echo is from the bright region, orbit number 10737, and was collected using the 4 MHz center frequency mode. The simulation that produced the pink echo included two CO2 layers separated by a dusty ice layer at the base of the SPLD (rightmost column of figure 1), and also used a 4 MHz center frequency. All three subsurface layers were 12 m thick. The second simulation (cyan dashed line) represents the "wet based" scenario proposed by Orosei et al. [2018] and included no subsurface layering (leftmost column of figure 1). A basal permittivity consistent with that of a brine was used for the bottom half-space.*

*Results and Discussion*

For each simulation we record the ratio between the basal (subsurface) and surface reflection power ($P_{ss}/P_s$), sometimes referred to as the normalized basal echo power. It is this quantity that previous studies measured directly from the data and then inverted to estimate basal permittivity under the assumption of a single simple boundary [2, 3, 4]. Table 1 summarizes our results. The median normalized basal echo power measured inside the bright region is between 0.08 – 2.52 dB depending on the MARSIS center frequency (dashed lines in figure 3). Note that while these values are broadly consistent with those previously reported [2], they may vary slightly due to differences in how the "bright region" was defined (see methods).

*Table 1: Summary of simulation results. Columns contain the maximum normalized basal echo power for each subsurface layering scenario. Results are shown for simulations including both a water ice surface and a CO2 ice surface (in parentheses).*

| Basal Layering Scenario | 3 MHz | 4 MHz | 5 MHz |
| --- | --- | --- | --- |
| No Basal Layering | -3.57 (-3.42) | -3.69 (-3.42) | -3.73 (-3.33) |
| Single CO2 Layer | 1.91 (2.05) | 1.81 (2.10) | 1.80 (2.23) |
| Two CO2 Layers and One Dusty Ice Layer | 4.44 (4.57) | 4.34 (4.65) | 4.57 (4.99) |
|  |  |  |  |

Despite the fact that the inclusion of a surface CO2 ice layer increased the normalized basal echo power, the effect was not large enough on its own to account for the enhanced reflection power observed by MARSIS in the bright region. This finding confirms a similar conclusion drawn in Orosei et al. [2]. Unlike that work, we found that a single CO2 ice layer at the base of the SPLD could produce a powerful reflection similar to those seen in the bright region in both the 4 and 5 MHz data (Figure 3, top). The primary difference between our respective simulations is the choice of basal temperature. Orosei et al. [2] chose a relatively warm 205 K, while our simulations were run assuming a basal temperature of 175 K (see methods), which is more in line with estimates made under typical Mars conditions [8] and may actually be an overestimation for the south polar region [9]. At lower temperatures, the overlying water ice absorbs less energy, resulting in more powerful reflections from the base.

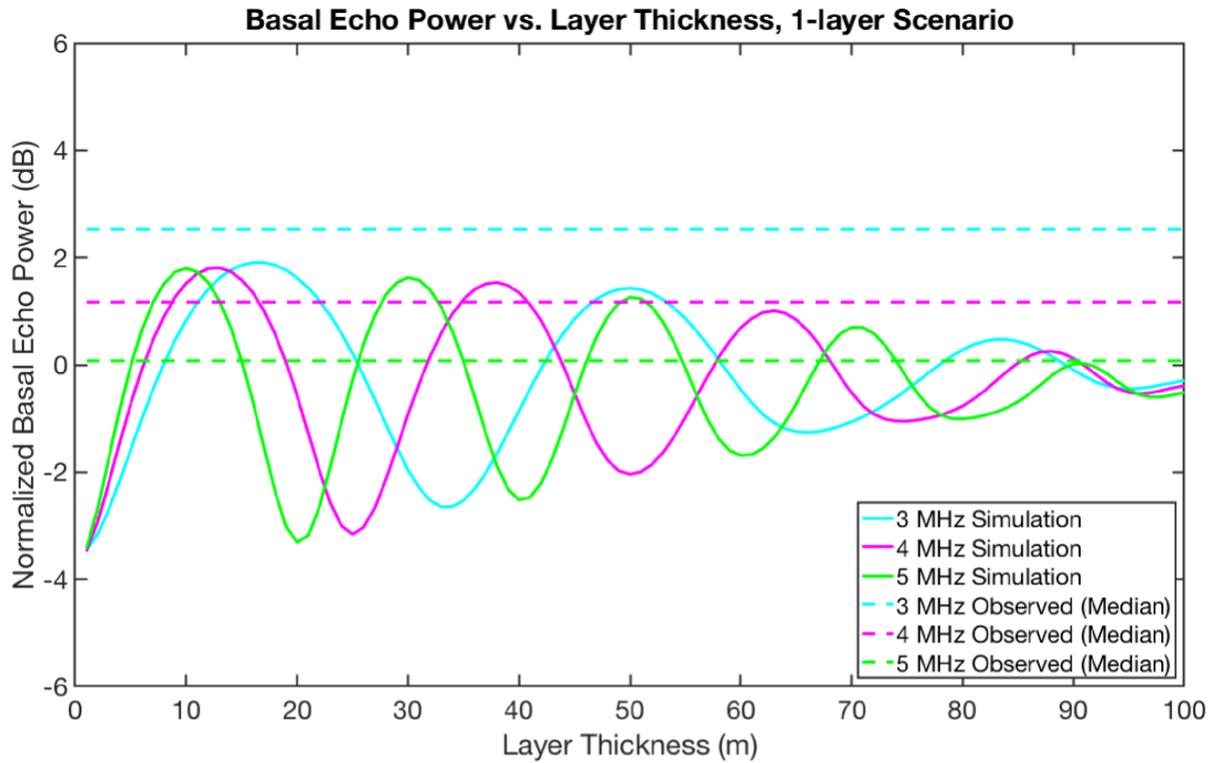
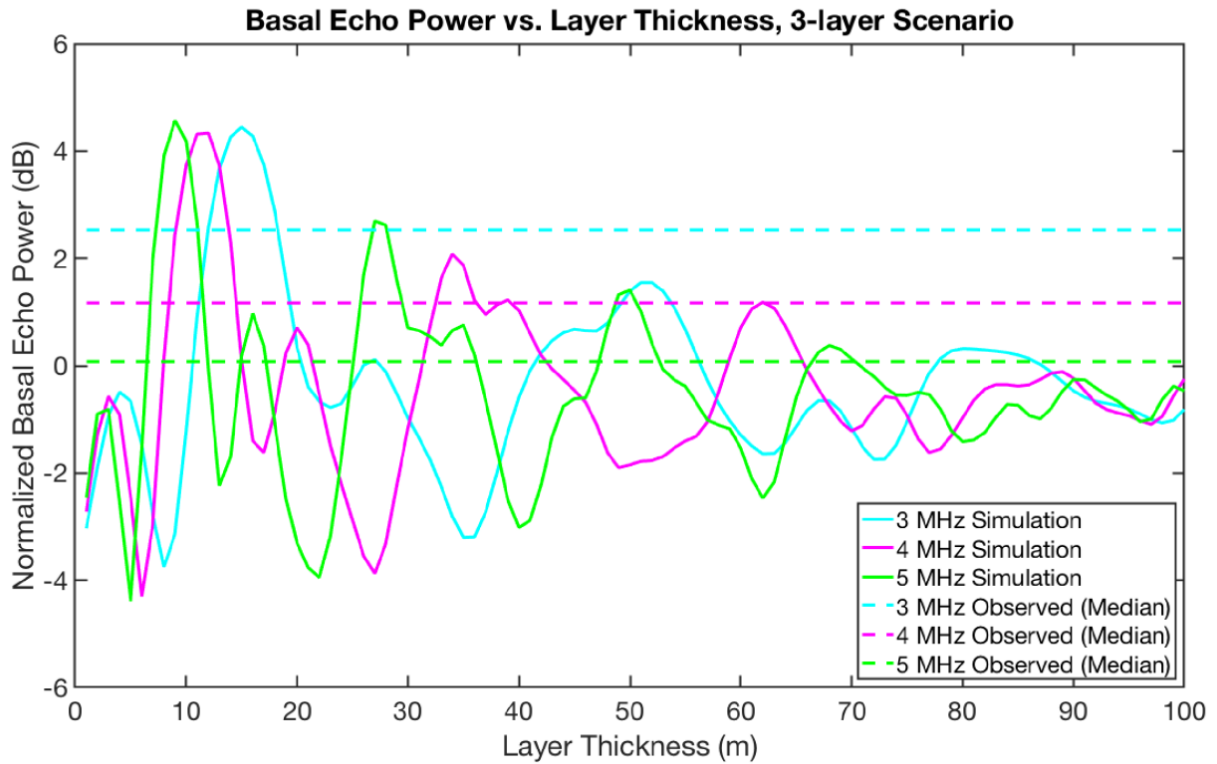

*Figure 3:* Simulated normalized basal echo power as a function of layer thickness. Solid lines are simulation results, dashed lines are the median values measured within the bright region at each frequency. **Top:** Single $CO_2$ ice layer scenario. **Bottom:** Two $CO_2$

*ice layers separated by a layer of dusty ice. Results are shown for simulations without surface CO2 ice.*

While a single CO2 ice layer can produce basal echoes consistent with most observed values, it cannot explain the brightest observed reflections. In contrast, our three-layer simulations were able to produce such reflections easily (Figure 3, bottom). For some layer thicknesses, we were able to produce basal echo powers over 4.5 dB. If we assume the basal reflection is caused by a single boundary between the SPLD ice and the underlying material (an erroneous assumption in this case) and apply a simple inversion method for retrieving basal permittivity [20], the strongest simulated echo powers produce permittivities in excess of 50. Normally, such a high permittivity is only observed in the presence of liquid water, but in this scenario none is actually present. The strong reflection is instead the result of constructive interference between multiple subsurface layers with much lower permittivities.

It is important to stress that our simulations are by no means comprehensive. The shape and thickness of the SPLD is such that much of the underlying material does not outcrop at troughs or scarps [15, 16], meaning any number of basal scenarios are possible. However, the layer thicknesses chosen for our simulations are in line with previous measurements of outcropping layers [15, 17], and large slabs of CO2 ice are known to be present in the SPLD [21, 22]. Recent work has also shown that the base of the North Polar Layered Deposits (NPLD) consists of alternating layers of sand and water ice, which may be remnants of older, more extensive polar caps [23]. It has been suggested that similar deposits may exist below the SPLD [24]. Therefore, while we cannot test every possible scenario, we can be confident that the scenarios we chose are plausible, and do not require us to invoke any unique subsurface phenomena or materials.

There is additional evidence supporting the interpretation of MARSIS echoes as the result of interference between multiple layers. Comparisons between outcrop stratigraphy and MARSIS data in one region of the SPLD found that MARSIS reflections correlated with packets of layers, rather than specific layer boundaries [25]. As stated before, interference between multiple layers is already believed to play a large role in determining radar echo powers in the NPLD [11, 12]. Previously measured SPLD layer thicknesses are consistent with the scales at which one would expect constructive and destructive interference in MARSIS data [15, 17]. Orosei et al. [2] reported that there are significant differences in observed basal reflection power in the bright region depending on which center frequency is used, a finding that we confirmed here (figure 3, dashed lines). Such differences cannot be easily explained under the assumption of a single boundary between the overlying ice and the bedrock, but are expected for scenarios with strong interference, which is a frequency-dependent phenomenon.

One potential weakness of our interpretation is that specific layer thicknesses and relatively uncommon (though still known to be present regionally) materials are required to produce the most powerful observed reflections. However, bright reflections are limited to small regions and are quite rare overall [2, 4]. It could be that the somewhat strict requirements for generating the most powerful constructive interference helps explain this localized behavior. The layers responsible for the basal reflection may only approach the requisite thickness in this location, or the preservation of ancient $CO_2$ ice may have been limited to small patches due to basal topography or localized climate conditions. It is also possible to create constructive interference through many other scenarios not explored in this paper.

The prospect of liquid water below the Martian ice cap is intriguing, but recent studies have questioned the feasibility of that interpretation [8, 9, 10]. Herein, we have demonstrated that observed MARSIS reflections can in fact be explained by phenomenon and materials already known to be present on modern day Mars. This new, independent interpretation is consistent with previous findings, and provides a plausible alternative to any hypothesis requiring liquid water.

## Methods

*MARSIS Data and Instrument Description*

The Mars Advanced Radar for Subsurface and Ionospheric Sounding (MARSIS) arrived at Mars in 2003 on board the Mars Express orbiter. MARSIS is a nadir-looking radar sounder capable of sampling both the ionosphere and the near-subsurface. MARSIS operates at four different center frequencies: 1.8 MHz, 3 MHz, 4 MHz, and 5 MHz, though the 1.8 MHz mode is usually reserved for ionospheric sounding and is rarely used to probe the subsurface. The transmitted signal linearly sweeps a 1 MHz bandwidth around the specified center frequency over 250 μs and is commonly referred to as a "chirp." The instrument transmits the chirp toward the surface through a 40 m dipole antenna, and then records the reflected signal as a function of time. The received signal is put through standard pulse-compression processing, meaning it is correlated with the complex conjugate of the transmitted chirp, resulting in a range (vertical) resolution of approximately 150 m in free space. The cross-track lateral resolution is typically between 10-30 km depending on spacecraft altitude. In standard modes, the along-track lateral resolution is reduced through on-board synthetic-aperture radar (SAR) processing to 5-10 km [26].

However, in this paper we have compared our results to the data provided by Orosei et al. [2], which was collected in the non-standard "superframe" mode. This allowed the authors to bypass on-board processing and instead process and analyze the raw data on the ground. SAR processing was not applied to this data, but due to the smoothness of the SPLD surface at length scales relevant to MARSIS, the lateral resolution can be expressed as the diameter of the Fresnel zone, 6-10 km depending on altitude and frequency [2].

In order to compare our simulation results to MARSIS observations, we calculate the median observed basal echo power at each frequency within the bright region. Our results (2.52 dB, 1.17 dB, and 0.08 dB at 3, 4, and 5 MHz respectively) are consistent with those shown on supplementary figures in Orosei et al. [2]. Any small variations are likely due to differences in how the bright region was defined. While Orosei et al. [2] mapped the outline of the bright region by hand, we instead selected every echo within a 10 km radius of the bright region's central point (193°E, 81.1°S). This resulted in an area consistent with the approximately 20 km by 30 km dimensions reported in previous work [2].

*Calculating Reflectivity*

The only significant difference between our simulations and those done previously is that we do not use the recursive method outlined by Ulaby et al. [27] to calculate reflectivity. Instead, we use the "matrix method" to calculate the reflectivity of a finite stack of layers with arbitrary thickness and permittivity (i.e. composition) sandwiched between two half-spaces. This choice has no significant impact on the final result, the matrix method was chosen solely for its relative simplicity and ease of implementation. While we refer to Born and Wolf [28] and Pascoe [29] for a full derivation, a simple description is provided here.

The matrix method assumes a plane wave incident on an infinite planar surface traveling through a lossless medium (air). This assumption is valid in the case where the surface is smooth at the wavelength scale, as is the case for MARSIS and the SPLD [2, 30]. In principle the matrix method can be used for a large range of incidence angles, but we assume normal incidence for MARSIS. Each layer in the model is described by a single 2x2 matrix whose elements are a function of the layer's complex permittivity and the radar frequency. Multiplying each of these matrices together, we arrive at a single 2x2 characteristic matrix representing the full stack of layers. The elements of the characteristic matrix are then combined using a standard set of formulas to calculate the total effective reflectivity and transmissivity at a single frequency, accounting for the effects of interference, multiple internal bounces, and power absorption. Following the method of Lalich et al. [13], we repeat this series of calculations for each frequency in the transmitted signal to construct the final effective reflectivity, which is used as the basis for our simulation as described in the main text.

*Layer Permittivity*

To calculate the total effective reflection coefficient, we must assign each model layer a complex relative permittivity. Layers are made up of four materials: atmosphere, water ice, CO2 ice, and basaltic rock or dust. We assign the layer representing the Martian atmosphere a permittivity of 1, equal to that of free space.

The complex permittivity of water ice depends on frequency and temperature. While ice temperature changes as a function of depth in the SPLD [8] the effect is weak for typical values of geothermal flux, which result in basal temperatures between approximately 165 and 180 K. Rather than split the SPLD water ice into multiple layers with slightly different permittivities, we calculated the complex permittivity of water ice at our chosen basal temperature (175 K) according to the formulas provided in Matzler [31], and applied the result ($3.11 - i*4.3 \times 10^{-7}$) to the entire depth of the SPLD. This simplification means that we are overestimating the water ice temperature for most of the deposit, which has the effect of increasing radar absorption and thus biasing our basal echo powers to lower values. Despite this negative bias, our simulations still produce reflections as bright or brighter than those observed by MARSIS.

In our model, the bulk of the SPLD is represented not by pure water ice, but by a mixture of water ice and basaltic dust. For the permittivity of the basaltic dust component we adopted the Shergottite value of Nunes and Phillips ($8.8 - i*0.017$) [14], which was also the value chosen for previous MARSIS studies of the SPLD [2; 3]. We then applied the Maxwell Garnett mixing formula [32] to determine the complex permittivity of water ice with 10% bulk dust content, consistent with previous estimates of SPLD composition ($3.48 - i*7 \times 10^{-4}$) [18]. The bedrock layer of our simulations was assigned a permittivity equal to that of pure Shergottite basalt 14].

For CO2 ice we chose a permittivity consistent with laboratory measurements ($2.2 - i*4.5 \times 10^{-4}$) [33]. While the real part of the permittivity was easily measured, Pettinelli et al. [33] were only able to place an upper bound on the imaginary part, which is primarily responsible for determining the amount of energy absorbed by the material. It is worth noting that the CO2 ice present in the upper SPLD appears to be remarkably low loss in SHARAD data [21], and therefore by adopting the maximum possible value for the imaginary part of the permittivity we may be overestimating the power absorbed by the CO2 layers. This would result in another small bias toward lower basal echo powers.